\documentclass[
    ,final            
epsfig  ]
  {aipproc}

\layoutstyle{6x9}

\begin{document}

\title{Sivers asymmetry for the proton and the neutron}

\classification{13.88.+e,12.39.-x,21.45.-v}
\keywords      {DIS, transversity, neutron structure}

\author{S. Scopetta}{
 address={
Dipartimento di Fisica, Universit\`a degli Studi
di Perugia,
via A. Pascoli,
06100 Perugia, Italy
},
 altaddress=
{INFN, sezione di Perugia, via A. Pascoli,
06100 Perugia, Italy
}
}

\author{A. Courtoy} {
address={
Departament de Fisica Te\`orica, Universitat de Val\`encia
\\
and Institut de Fisica Corpuscular, Consejo Superior de Investigaciones
Cient\'{\i}ficas
\\
46100 Burjassot
(Val\`encia), Spain
}
}

\author{F. Fratini}{
address={
Dipartimento di Fisica, Universit\`a degli Studi
di Perugia,
via A. Pascoli,
06100 Perugia, Italy
},
}

\author{V. Vento}{
address= {
Departament de Fisica Te\`orica, Universitat de Val\`encia
\\
and Institut de Fisica Corpuscular, Consejo Superior de Investigaciones
Cient\'{\i}ficas
\\
46100 Burjassot
(Val\`encia), Spain
},
altaddress={
TH-Division, PH Department, CERN, CH-1211 Gen\`eve 23, 
Switzerland
}
}

\begin{abstract}

A formalism is presented to evaluate the Sivers function
in constituent quark models.
A non-relativistic reduction of the
scheme is performed and applied to the Isgur-Karl model. 
The results obtained are consistent with a sizable Sivers
effect and the signs for the $u$ and $d$ flavor contributions turn
out to be opposite. 
The Burkardt Sum Rule is fulfilled to a large extent.
After the estimate of the QCD evolution of
the results from the momentum scale of the model to the experimental
one, a reasonable agreement with the available data
is obtained.
A calculation of nuclear effects in the extraction of neutron
single spin asymmetries in semi-inclusive deep inelastic scattering
off $^3$He is also described. In the kinematics of forth-coming
experiments at JLab, it is found that the nuclear effects arising within an 
Impulse Approximation approach are under control.
\end{abstract}

\maketitle


\section{The Sivers function in Constituent Quark Models}

The partonic structure of transversely polarized nucleons
is one of their less known features
(for a review, see, e.g.,
Ref. \cite{bdr}).
The work presented here aims to contribute to the effort
of shedding some light on it.

Semi-inclusive deep inelastic scattering (SIDIS)
is one of the proposed
processes to access the parton distributions (PDs)
of transversely polarized hadrons.
SIDIS of unpolarized electrons off a transversely polarized target
shows azimuthal asymmetries,
the so called ``single spin asymmetries'' (SSAs) \cite{Collins}.
The SSAs are 
due to two physical mechanisms,
whose contributions can be distinguished
\cite{mu-ta,ko-mu,boer,bac1}.
One of them is the Collins mechanism, due to parton final state interactions
(FSI) in the production of a hadron
by a transversely polarized quark
\cite{Collins}.
The other is the Sivers mechanism \cite{sivers},
producing
a term in the SSA which is given by the product of
the unpolarized fragmentation function with
the Sivers PD,
describing the number density of unpolarized quarks
in a transversely polarized target.
The Sivers function is a time-reversal odd,
Transverse Momentum Dependent (TMD) PD;
for this reason it was believed to vanish due to time reversal invariance.
However, this argument was invalidated by a calculation
in a spectator model \cite{brohs}, following the observation
of the existence of leading-twist
Final State Interactions (FSI) \cite{brodhoy}.
The current wisdom is that a non-vanishing Sivers function
is generated by the gauge link in the definition of TMDs
\cite{coll2,jiyu,bjy,adra}, whose
contribution does not vanish in the light-cone gauge,
as happens for the standard PD functions.
Recently, the first data
of SIDIS off transversely polarized targets have been published,
for the proton \cite{hermes} and the deuteron \cite{compass}.
It has been found that, while the Sivers effect
is sizable for the proton, it becomes negligible for the deuteron,
so that apparently the neutron contribution cancels the proton one,
showing a strong flavor dependence of the mechanism.
Different parameterizations of the available
SIDIS data have been published \cite{ans,coll3,Vogelsang:2005cs}, 
still with large error bars.
New data, which will
reduce the uncertainties on the extracted Sivers function
and will help discriminate between different theoretical predictions,
will be available soon. 

This experimental scenario motivates theoretical
estimates of this quantity. Since a calculation from first
principles in QCD is not yet possible, 
several model evaluations exist, in a quark-diquark model
\cite{brohs,jiyu,bacch}; in the MIT bag model \cite{yuan}; 
in a light-cone model \cite{luma}; in a nuclear
framework, relevant to proton-proton collisions \cite{bianc}.
\begin{figure}
  \includegraphics[height=.2\textheight]{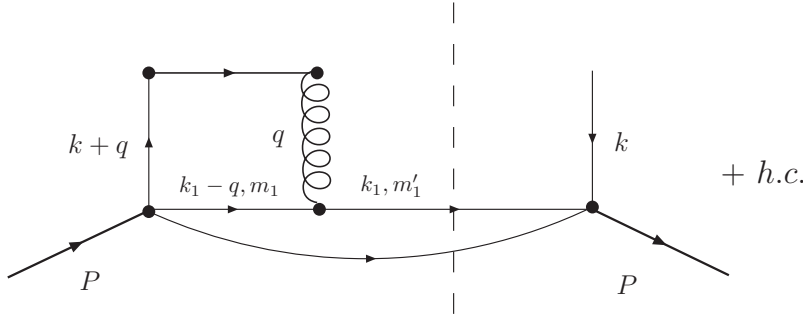}
  \caption{The contributions to the Sivers
function in the present approach.}
\end{figure}
In here, a Constituent Quark Model (CQM) calculation will be described
\cite{nostro}.
CQM calculations of PDs are based on a two steps procedure
(see, e.g., Ref. \cite{trvv}). First, the matrix element of the
proper operator is evaluated using the wave functions of the model;
then, a low momentum scale is ascribed to
the model calculation and QCD evolution is used
to evolve the observable calculated in this low energy scale to
the high momentum one, where DIS experiments are carried out
(see, e.g., Ref \cite{trvv}). Such
procedure has proven successful in describing the gross features of
PDs (see, e.g., \cite{h1,oam}) and GPDs (see, e.g.
\cite{epj,bpt}), by using different CQMs. Similar expectations
motivate the present study of the Sivers function.

The formalism of Ref. \cite{nostro} to calculate the valence quark
contribution to the Sivers function from any CQM is summarized here.
A difference in the calculation of TMDs, with respect to calculations
of PDs and GPDs, is that the leading twist contribution to the
FSI has to be evaluated.
All the technical steps of our procedure can be found in Ref. \cite{nostro},
where a workable formula is derived. The main approximations
have been: i) only the valence quark sector is investigated; ii) the leading 
twist FSI are taken into account at leading order, i.e.
only one-gluon-exchange (OGE) FSI has been evaluated in the
treatment of the gauge link (see Fig. 1); 
iii) the resulting interaction has been obtained
through a non-relativistic (NR) reduction of the relevant operator,
according to the philosophy of constituent quark models \cite{ruju},
leading to a potential $V_{NR}$.
In terms of the latter, the Sivers function for the
quark of flavor ${\cal Q}$,
$f_{1T}^{\perp {\cal Q}} (x, {k_T} )$, can be written
(cf. Fig. 1 for the labels of the momenta and helicities):
\begin{eqnarray}
f_{1T}^{\perp {\cal Q}} (x, {k_T} ) & = &
\Im
\left \{
- i g^2
{
M^2 \over k_x
}
\int
d \vec k_1
d \vec k_3
{d^2 \vec q_T \over (2 \pi)^2}
\delta(k_3^+ - xP^+)
\delta(\vec k_{3 T} + \vec q_T - \vec k_T) {\cal M}^{\cal Q}
\right \}
\label{start2}
\end{eqnarray}
where $g$ is the strong coupling constant, $M$ the proton mass,
and
\begin{eqnarray}
{\cal M}^{u(d)} & = &
\sum_{m_1,m_1',m_3,m_3'}
\Phi_{sf,S_z=1}^{\dagger}
\left ( \vec k_3, m_3; \vec k_1, m_1;
\, \vec P - \vec k_3 - \vec k_1,  m_n  \right )
\nonumber
\\
& \times &
{ 1 \pm \tau_3(3) \over 2 }
V_{NR}(\vec k_1, \vec k_3, \vec q)
\nonumber
\\
& \times &
\Phi_{sf , S_z=-1}
\left (\vec k_3 + \vec q, m_3'; \, \vec k_1 -
\vec q, m_1';
\, \vec P - \vec k_3 - \vec k_1,  m_n  \right )~.
\label{Mu}
\end{eqnarray}

Eq. (\ref{start2}), with ${\cal{M}}^{u(d)}$
given by Eq. (\ref{Mu}),
provides us with a suitable formula
to evaluate the Sivers function,
once the spin-flavor wave function of the proton
in momentum space, i.e. the quantity
$\Phi_{sf}$, is available
in a given constituent quark model.
A few remarks are in order. First of all, the helicity conserving
part of the global interaction
does not contribute to the Sivers function.
Besides, one should notice that, in an extreme NR limit, the Sivers function
would turn out to be identically zero.
In our approach, it is precisely the interference of the small
and large components in the four-spinors
of the free quark states which leads to a non-vanishing
Sivers function, even from
the component with $l = 0$ of the target wave function.
Effectively, these interference terms in the interaction are
the ones that, in other approaches, arise due to the wave function
(see, e.g., the MIT bag model calculation \cite{yuan}).

There are many good reasons to use the Isgur-Karl model \cite{ik} to test 
the performance of the approach. 
First of all, the IK is the typical CQM,
successful in reproducing the low-energy properties of the nucleon.
In particular, in the IK model, 
one expects small corrections from terms $O\left ( { k^2 / m^2} \right )$.
Besides, one of the features of the IK model is that the OGE
mechanism \cite{ruju}, which reduces the degeneracy of the spectrum,
is taken into account. It is therefore natural
to study our formalism, based on OGE FSI,
within the IK framework.
Concerning PDs,
it has been shown that the IK model can describe their gross
features, once QCD evolution of the proper matrix elements
of the corresponding twist-2 operators is performed from the scale
of the model to the experimental one \cite{trvv,h1,oam}.
Reasonable predictions of GPDs have also been obtained \cite{epj},
and this makes particularly interesting the evaluation of the Sivers
function in the IK model.
The final expressions of the Sivers function in the IK model
are rather involved and not presented here, since they can be found
in Ref. \cite{nostro}.

To evaluate numerically Eq. (\ref{start2}), the
strong coupling constant $g$, and therefore
$\alpha_s(Q^2)$, has to be fixed.
Here, the prescription introduced in the past for calculations of
PDFs in quark models (see, i.e., Ref. \cite{trvv}) will be used.
It consists in fixing the momentum scale of the model,
the so-called hadronic scale $\mu_0^2$, according to the
amount of momentum carried by the valence quarks in the model.
In the approach under scrutiny, only valence quarks contribute.
Assuming that all the gluons and sea pairs in the proton
are produced perturbatively according to NLO evolution equations,
in order to have $\simeq 55 \% $ of the momentum
carried by the valence quarks at a scale of 0.34 GeV$^2$,
as in
typical low-energy parameterizations, one finds,
that $\mu_0^2 \simeq 0.1$ GeV$^2$
if $\Lambda_{QCD}^{NLO} \simeq 0.24$ GeV.
This yields $\alpha_s(\mu_0^2)/(4 \pi) \simeq 0.13$ \cite{trvv}.
For an easy presentation,
the quantity which is usually shown for the results of calculations
or for data of the Sivers function is its first moment, defined
as follows :
\begin{equation}
f_{1T}^{\perp (1) {\cal Q} } (x)
= \int {d^2 \vec k_T}  { k_T^2 \over 2 M^2}
f_{1T}^{\perp {\cal Q}} (x, {k_T} )~.
\label{momf}
\end{equation}
\begin{figure}
\includegraphics[width=.49\textwidth]{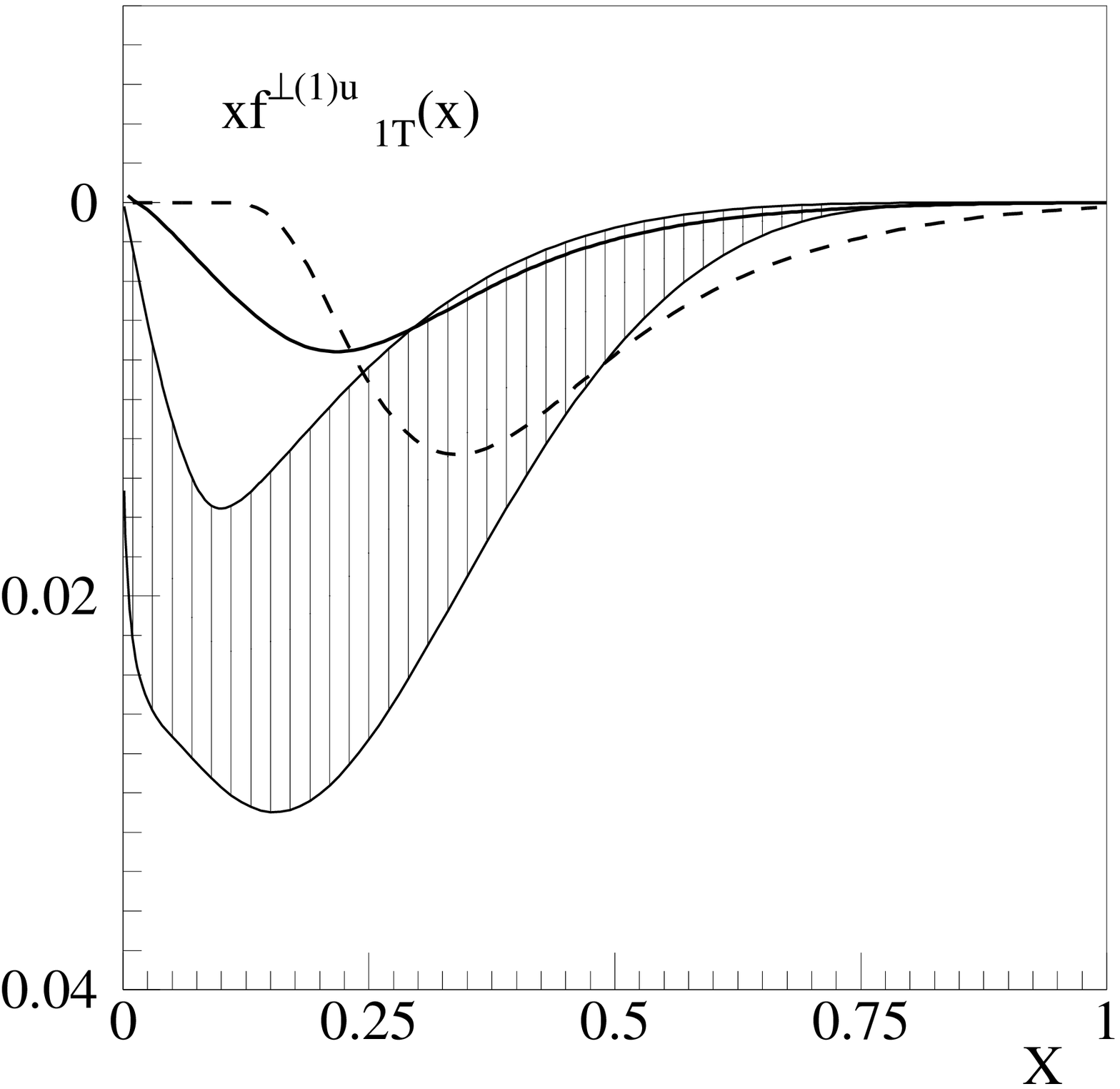}
\includegraphics[width=.49\textwidth]{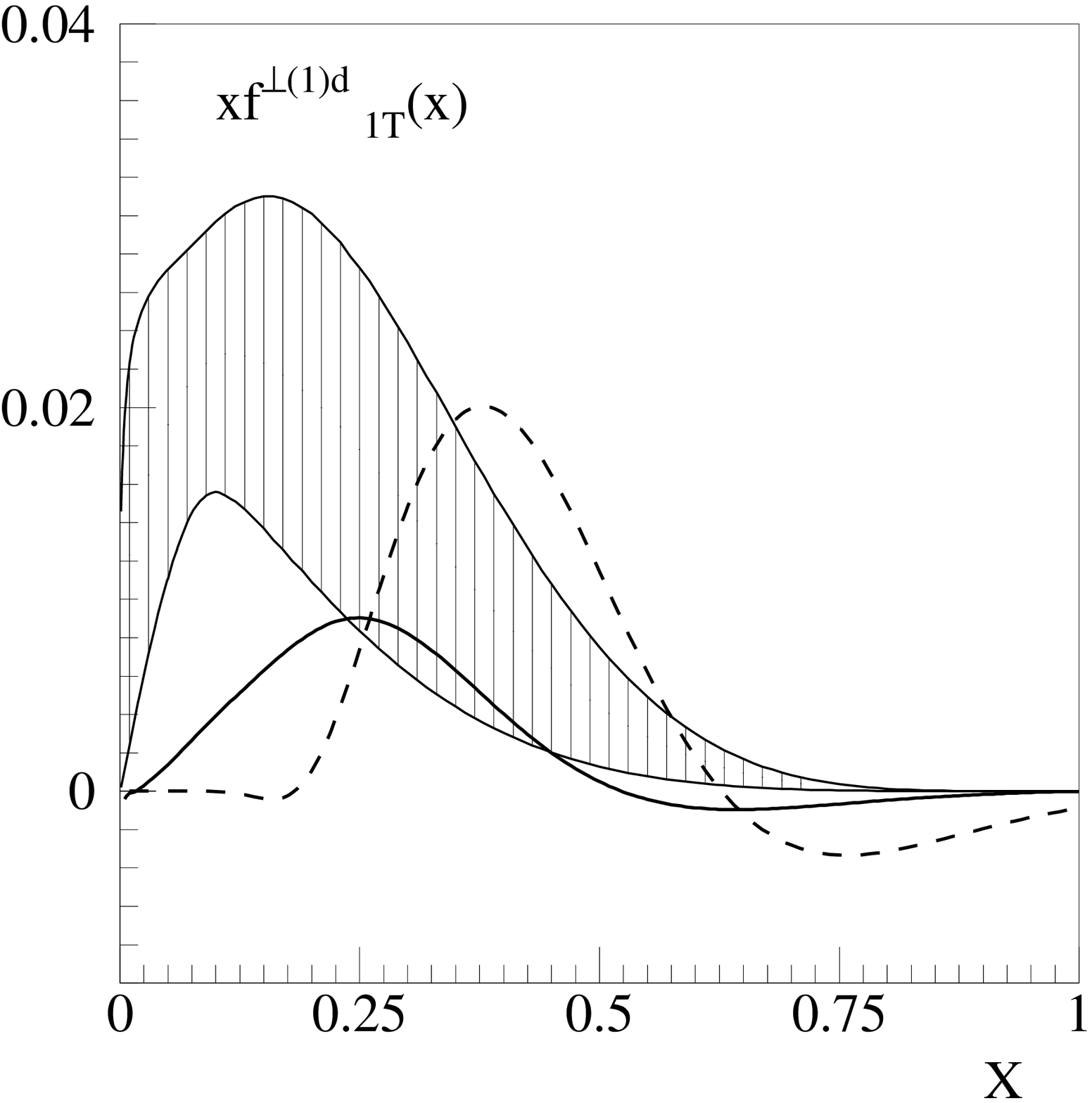}
\caption{
Left (right): the quantity $f_{1T}^{\perp (1)q }(x) $, Eq. (\ref{momf}),
for the $u$ ($d$) flavor.
The dashed curve is the result of the present approach
at the hadronic scale $\mu_0^2$.
The full curve represents the evolved distribution after
standard NLO evolution (see text).
The patterned area represents the $1 - \sigma$ range of the best fit 
of the HERMES data
proposed in Ref. \cite{coll3}.
}
\end{figure}

The results of the present approach
for the moments Eq. (\ref{momf}) are given
by the dashed curves in Fig. 2 for
the flavors $u$ and $d$.
They are compared
with a parameterization of the HERMES data,
corresponding to an experimental scale of $Q^2=2.5$ GeV$^2$
\cite{coll3}.
The patterned area represents the $1-\sigma$ range
of the best fit proposed in Ref. \cite{coll3}.
Clearly, a different sign for the $u$ and $d$ flavor is found.

Let us see now how the results of the calculation
compare with the Burkardt sum rule
\cite{Burkardt:2004ur}, 
which follows from general principles
and must be satisfied at any scale.
If the proton is polarized in the positive $y$ direction,
in our case, where only valence quarks are present,
the Burkardt sum rule reads:
\begin{equation}
\sum_{{\cal Q}=u,d} \langle k_x^{\cal{Q}} \rangle = 0~,
\label{burksr}
\end{equation}
where
\begin{equation}
\langle k_x^{\cal{Q}} \rangle = - \int_0^1 d x \int d \vec k_T
{k_x^2 \over M}  f_{1T}^{\perp \cal{Q}} (x, {k_T} )~.
\label{burs}
\end{equation}
Within our scheme, at the scale of the model, it is found
$\langle k_x^{u} \rangle = 10.85$ MeV,
$\langle k_x^{d} \rangle = - 11.25$ MeV and, 
in order to have an estimate
of the quality of the agreement of our results with
the sum rule, we define the ratio
\begin{equation}
r= {
\langle k_x^{d} \rangle+
\langle k_x^{u}\rangle
\over
\langle k_x^{d} \rangle-
\langle k_x^{u} \rangle}~,
\label{rbur}
\end{equation}
obtaining $r \simeq 0.02$, so that we can say that our calculation
fulfills the Burkardt sum rule to a precision of a few percent.

The magnitude of the results
is close to that of the data,
although
they have a different shape: the maximum (minimum)
is predicted at larger values of $x$.
One should anyway realize that one step of the analysis
is still missing: the scale of the model,
$\mu_0^2$, is much lower than the one of the
data, which is $Q^2 =2.5$ GeV$^2$.
For a proper comparison, the QCD
evolution from the model scale to the experimental one
would be necessary. 
Unfortunately, the Sivers function is a TMD PDs and the evolution
of this class of functions is, to a large
extent, unknown.
In order to have an indication of the effect of the
evolution, we perform a NLO evolution of the model
results assuming, for the moments of the Sivers function,
the ones defined in Eq. (\ref{momf}),
the same anomalous dimensions of the unpolarized PDFs.
As described in the previous section,
the parameters of the evolution have been fixed
in order to have a fraction $\simeq 0.55$ of the momentum
carried by the valence quarks at 0.34 GeV$^2$, as in
typical parameterizations of PDFs, starting from
a scale of $\mu_0^2 \simeq 0.1$ GeV$^2$ with only valence quarks.
The final result is given by the full curve in Fig. 2 
for the $u$ and $d$ flavor.
As it is clearly seen, the agreement with data
improves dramatically and their
trend is reasonably reproduced at least for $x \ge 0.2$.
Of course a word of caution is in order:
the performed evolution is not really correct.
In any case, an indication of two very important things
is obtained:
i) The evolution of the model result is necessary to estimate
the quantities at the momentum scale of experiments,
as it happens for standard PDs \cite{trvv,h1,oam};
ii) after evolution, the present calculation could be
consistent with data,
at least with the present ones, still
affected by large statistical and systematic errors.
One should notice that the agreement which is found
is better than that found in other model calculations
\cite{bacch,yuan},
especially for what concerns the fulfillment of the
Burkardt Sum Rule.

\section{The Sivers function from neutron
($^3$He) targets}

As we have discussed in the previous section,
the experimental scenario which arises from the analysis
of SIDIS off transversely polarized
proton and deuteron targets \cite{hermes,compass} is rather
puzzling. The data show
a strong, unexpected flavor dependence in the azimuthal
distribution of the produced pions. 
With the aim at extracting the neutron information
to shed some light on the problem,
a measurement of SIDIS
off transversely polarized $^3$He has been addressed \cite{bro},
and two experiments, planned to measure azimuthal asymmetries
in the production of leading $\pi^\pm$  from transversely
polarized $^3$He, are forth-coming at JLab \cite{ceb}.
Here, a recent, realistic analysis of SIDIS
off transversely polarized $^3$He \cite{mio} is described.
The formal expressions
of the
Collins and Sivers contributions to the azimuthal
Single Spin Asymmetry (SSA) for the production
of leading pions have been derived, in impulse approximation (IA),
including also the initial transverse momentum of the struck quark.
The final equations are rather involved and they are not
reported here. They can be found in \cite{mio}.
The same quantities have been then evaluated
in the kinematics of the planned 
JLab experiments.
Wave functions \cite{pisa} obtained within
the AV18 interaction \cite{av18} have been used for a realistic
description of the nuclear dynamics,
using overlap integrals evaluated in Ref. \cite{over},
and the nucleon structure has been described
by proper parameterizations of data or suitable model calculations
\cite{ans,model}.
The crucial issue of extracting
the neutron information from $^3$He data
will be now thoroughly discussed. 
As a matter of facts,
a model independent procedure, based
on the realistic evaluation
of the proton and neutron polarizations in $^3$He
\cite{old}, called respectively $p_p$ and $p_n$ in the following,
is widely used in inclusive 
DIS to take into account effectively  
the momentum and energy distributions
of the bound nucleons in $^3$He.
It is found that the same extraction technique
can be applied also in the 
kinematics of the proposed experiments, although
fragmentation functions, not only parton
distributions, are involved, as it can be seen
in Figs. 1 and 2. In these figures,
the free neutron asymmetry used as a model in the
calculation, given by a full line, is compared with two other quantities.
One is:
\begin{equation}
\bar A^i_n \simeq {1 \over d_n} A^{exp,i}_3~,
\label{extr-1}
\end{equation}
where $i$ stands for ``Collins'' or ``Sivers'',
$A^{exp,i}_3$ is the result of the full calculation, 
simulating data, and $d_n$ is the neutron dilution
factor. The latter quantity is defined as follows, for a neutron $n$
(proton $p$) in $^3$He:
\begin{eqnarray}
d_{n(p)}(x,z)=
{\sum_q e_q^2
f^{q,{n(p)}} 
\left ( x \right )
D^{q,h} \left ( z  \right )
\over
\sum_{N=p,n}
\sum_q e_q^2
f^{q,N} 
( x )
D^{q,h} 
\left ( z \right )
}
\label{dilut}
\end{eqnarray}
and, depending on the standard parton
distributions, $ f^{q,N} ( x )$,
and fragmentation functions, $D^{q,h} 
\left ( z \right )$,
is experimentally known (see \cite{mio} for details). 
$\bar A^i_n $ is given by the dotted curve in the figures.
The third curve, the dashed one, is given by 
\begin{equation}
A^i_n \simeq {1 \over p_n d_n} \left ( A^{exp,i}_3 - 2 p_p d_p
A^{exp,i}_p \right )~,
\label{extr}
\end{equation}
i.e. $^3$He is treated as a nucleus
where the effects of its complicated
spin structure, leading to a depolarization
of the bound neutron, together with the ones of
Fermi motion and binding, can be taken care
of by parameterizing the nucleon effective polarizations,
$p_p$ and $p_n$.
\begin{figure}
\includegraphics[width=.49\textwidth]{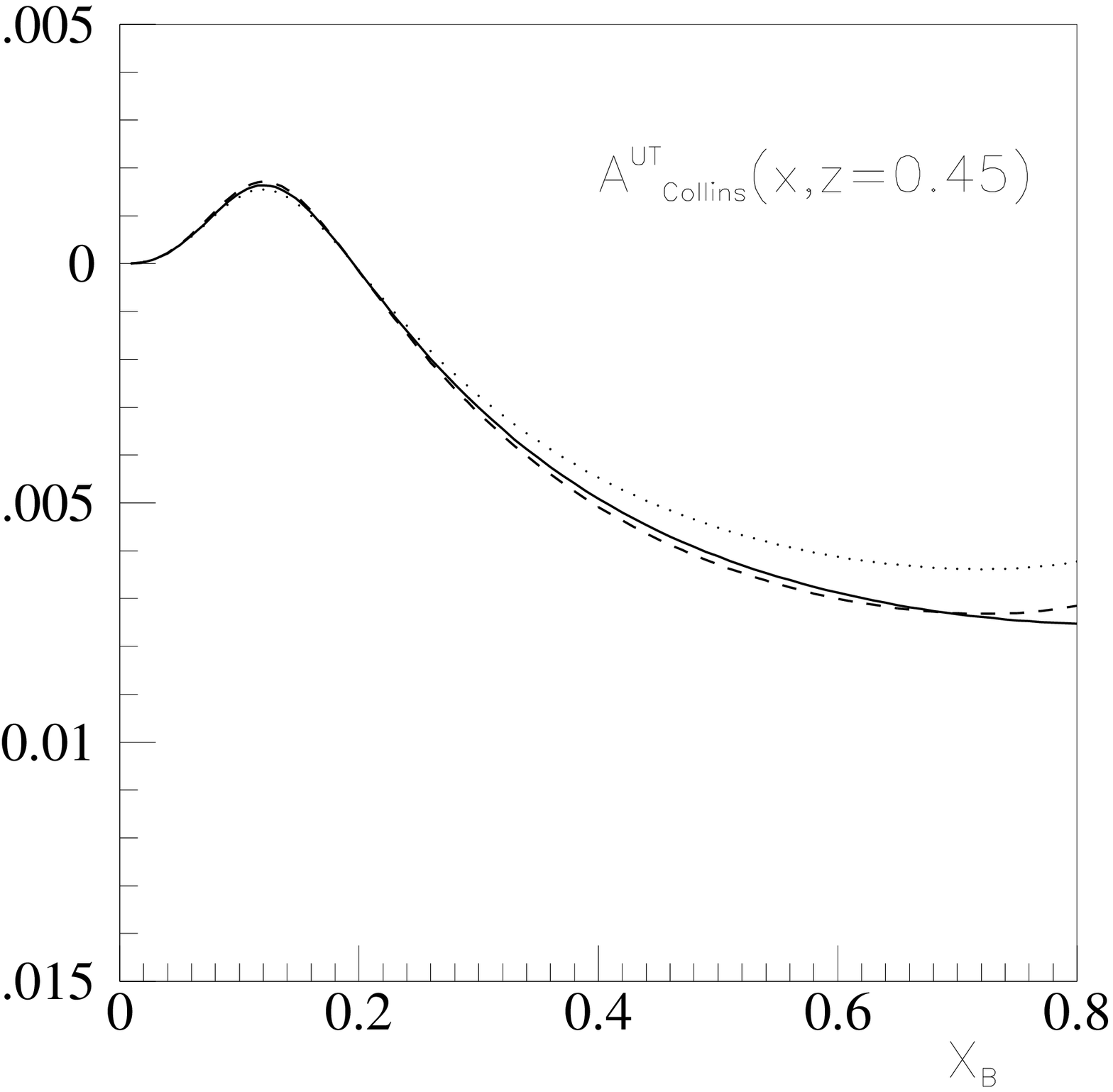}
\includegraphics[width=.49\textwidth]{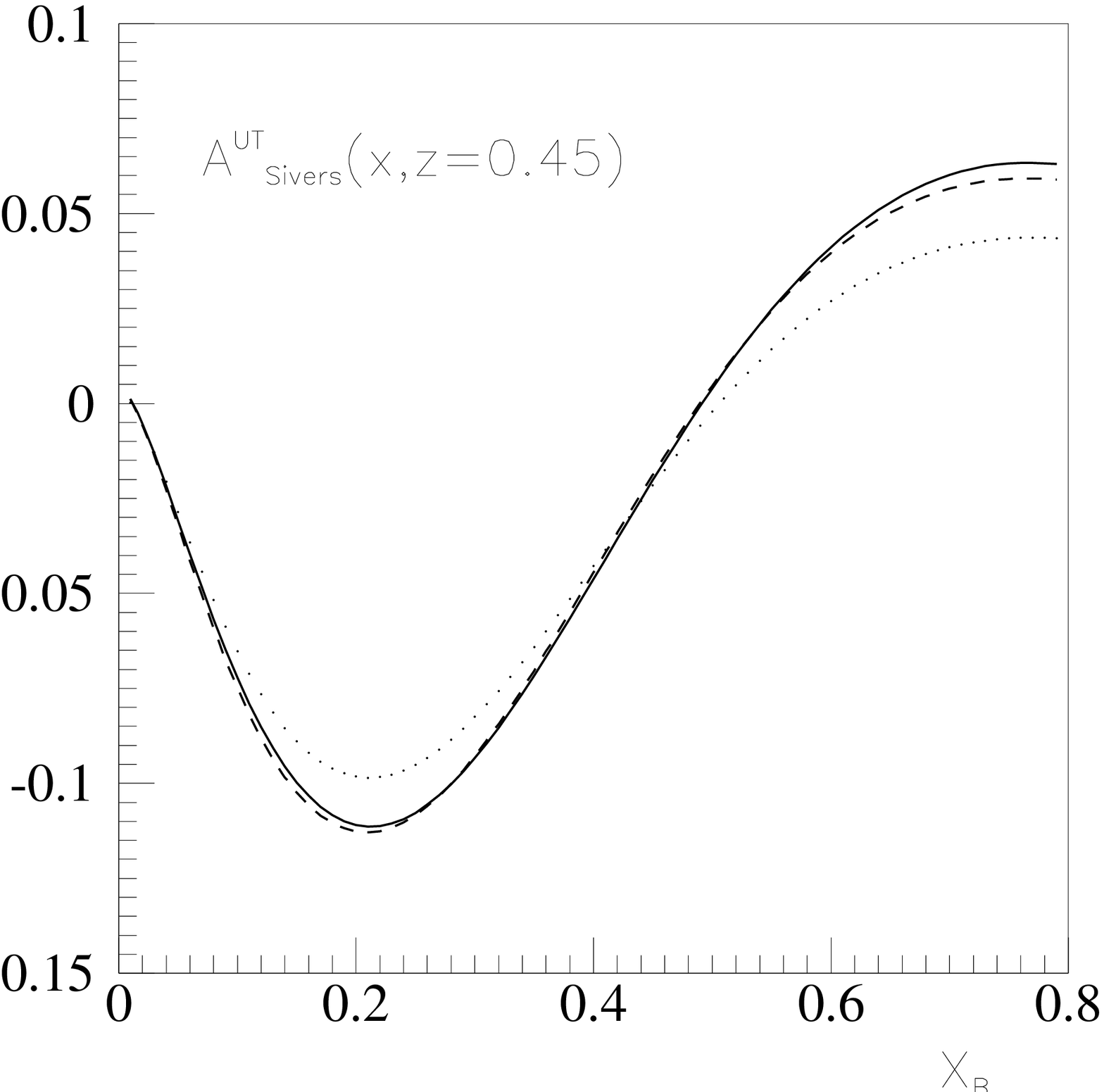}
\caption{Left (right)
The model neutron Collins (Sivers) asymmetry for
$\pi^-$ production
(full) in JLab kinematics, and the one extracted
from the full calculation taking into account
the proton effective polarization
(dashed), or neglecting it (dotted). 
The results are shown for {$z$}=0.45 and
$Q^2= 2.2$ GeV$^2$, typical values
in the kinematics of the JLab experiments.}
\end{figure}
One should realize that
Eq. (\ref{extr-1}) is the relation which should hold
between the $^3$He and the neutron SSAs if there were no nuclear effects,
i.e. the $^3$He nucleus were a system of free nucleons in a pure $S$ wave.
In fact, Eq. (\ref{extr-1}) can  be obtained from Eq. (\ref{extr}) by 
imposing $p_n=1$ and $p_p=0$.
It is clear from the figures that the difference 
between the full and dotted curves,
showing the amount of nuclear effects, is sizable,
being around 10 - 15 \% for any experimentally relevant $x$ and $z$,
while the difference between the dashed
and full curves reduces drastically
to a few percent, showing that the extraction
scheme Eq. (\ref{extr}) takes safely into account
the spin structure of $^3$He, together with Fermi
motion and binding effects. 
This important result is due to the peculiar kinematics
of the JLab experiments, which helps in two ways.
First of all, to favor pions from current fragmentation, 
$z$ has been chosen in the range $0.45 \leq z \leq 0.6$,
which means that only high-energy pions are observed.
Secondly, the pions are detected in a narrow cone around the direction
of the momentum transfer. As it is explained in \cite{mio},
this makes nuclear effects in the fragmentation 
functions rather small. The leading nuclear effects are then 
the ones affecting the parton distributions, already found
in inclusive DIS, and can be taken into account
in the usual way, i.e., using Eq. (\ref{extr}) for the extraction of the
neutron information. In the figures,
one should not take the 
shape and size of the asymmetries too seriously,
being the obtained quantities 
strongly dependent on the models chosen for the unknown distributions
\cite{model}.
One should instead consider the difference between
the curves, a model independent
feature which is the most relevant outcome of the present
investigation. 
The main conclusion is that Eq. (\ref{extr}) will be a valuable tool
for the data analysis of
the experiments \cite{ceb}.

The evaluation of possible effects beyond IA, such as 
final state interactions, and the inclusion in the scheme
of more realistic models of the nucleon structure, able to
predict reasonable figures for the experiments, are in progress.


\begin{theacknowledgments}
It is a pleasure to thank Alessandro Drago and Santiago Noguera
for interesting conversations.
This work is supported in part by the INFN-CICYT agreement,
by the Generalitat Valenciana under the contract
AINV06/118; by the Sixth Framework Program of the
European Commission under the Contract No. 506078 (I3 Hadron Physics);
by the MEC (Spain) under the Contract FPA 2007-65748-C02-01 and
the grants AP2005-5331 and PR2007-0048.
\end{theacknowledgments}



\bibliographystyle{aipprocl} 


\begin{thebibliography}{9}

\bibitem{bdr}
  V.~Barone, A.~Drago and P.~G.~Ratcliffe,
  Phys.\ Rept.\  {\bf 359} (2002) 1.

\bibitem{Collins}
  J.~C.~Collins,
  Nucl.\ Phys.\  B {\bf 396}, 161 (1993).

\bibitem{mu-ta}
  P.~J.~Mulders and R.~D.~Tangerman,
  Nucl.\ Phys.\  B {\bf 461} (1996) 197
  [Erratum-ibid.\  B {\bf 484} (1997) 538].

\bibitem{ko-mu}
  A.~M.~Kotzinian and P.~J.~Mulders,
  Phys.\ Lett.\  B {\bf 406} (1997) 373.

\bibitem{boer}
  D.~Boer and P.~J.~Mulders,
  Phys.\ Rev.\  D {\bf 57} (1998) 5780.

\bibitem{bac1}
  A.~Bacchetta, M.~Diehl, K.~Goeke, A.~Metz, P.~J.~Mulders and M.~Schlegel,
  JHEP {\bf 0702}, 093 (2007).

\bibitem{sivers}
  D.~W.~Sivers,
  Phys.\ Rev.\  D {\bf 41}, 83 (1990),
  Phys.\ Rev.\  D {\bf 43}, 261 (1991).
  
\bibitem{brohs}
  S.~J.~Brodsky, D.~S.~Hwang and I.~Schmidt,
  Phys.\ Lett.\  B {\bf 530}, 99 (2002).

\bibitem{brodhoy}
  S.~J.~Brodsky, P.~Hoyer, N.~Marchal, S.~Peigne and F.~Sannino,
  Phys.\ Rev.\  D {\bf 65}, 114025 (2002).

\bibitem{coll2}
  J.~C.~Collins,
  Phys.\ Lett.\  B {\bf 536}, 43 (2002).

\bibitem{jiyu}
  X.~d.~Ji and F.~Yuan,
  Phys.\ Lett.\  B {\bf 543}, 66 (2002).

\bibitem{bjy}
  A.~V.~Belitsky, X.~Ji and F.~Yuan,
  Nucl.\ Phys.\  B {\bf 656}, 165 (2003).

\bibitem{adra}
A. Drago, Phys. Rev. D {\bf 71}, 057501 (2005). 

\bibitem{hermes}
  A.~Airapetian {\it et al.}  [HERMES Collaboration],
  Phys.\ Rev.\ Lett.\  {\bf 94}, 012002 (2005).


\bibitem{compass}
  V.~Y.~Alexakhin {\it et al.}  [COMPASS Collaboration],
  Phys.\ Rev.\ Lett.\  {\bf 94}, 202002 (2005).


\bibitem{ans}
M.~Anselmino, M.~Boglione, U.~D'Alesio, A.~Kotzinian, F.~Murgia and A.~Prokudin,
  Phys.\ Rev.\  D {\bf 71}, 074006 (2005),
  M.~Anselmino, M.~Boglione, U.~D'Alesio, A.~Kotzinian, F.~Murgia and A.~Prokudin,
  Phys.\ Rev.\  D {\bf 72}, 094007 (2005).
  [Erratum-ibid.\  D {\bf 72}, 099903 (2005)]

\bibitem{coll3}
  A.~V.~Efremov, K.~Goeke, S.~Menzel, A.~Metz and P.~Schweitzer,
  Phys.\ Lett.\  B {\bf 612}, 233 (2005);
 J.~C.~Collins, A.~V.~Efremov, K.~Goeke, S.~Menzel, A.~Metz and P.~Schweitzer,
  Phys.\ Rev.\  D {\bf 73}, 014021 (2006).


\bibitem{Vogelsang:2005cs}
  W.~Vogelsang and F.~Yuan,
  Phys.\ Rev.\  D {\bf 72}, 054028 (2005).

\bibitem{bacch}
  A.~Bacchetta, A.~Schaefer and J.~J.~Yang,
  Phys.\ Lett.\  B {\bf 578}, 109 (2004);
A. Bacchetta, F. Conti, M. Radici,
arXiv:0807.0323 [hep-ph].

\bibitem{yuan}
  F.~Yuan,
  Phys.\ Lett.\  B {\bf 575}, 45 (2003);
  I.~O.~Cherednikov, U.~D'Alesio, N.~I.~Kochelev and F.~Murgia,
  Phys.\ Lett.\  B {\bf 642}, 39 (2006).


\bibitem{luma}
  Z.~Lu and B.~Q.~Ma,
  Nucl.\ Phys.\  A {\bf 741}, 200 (2004).


\bibitem{bianc}
  A.~Bianconi,
  arXiv:hep-ph/0702186.

\bibitem{nostro}
A. Courtoy, F. Fratini, S. Scopetta, V. Vento,
Phys. Rev. D (2008), in press;  arXiv:0801.4347 [hep-ph].


\bibitem{trvv}
  M.~Traini, A.~Mair, A.~Zambarda and V.~Vento,
  Nucl.\ Phys.\  A {\bf 614}, 472 (1997).

\bibitem{h1}
  S.~Scopetta and V.~Vento,
  Phys.\ Lett.\  B {\bf 424}, 25 (1998).


\bibitem{oam}
  S.~Scopetta and V.~Vento,
  Phys.\ Lett.\  B {\bf 460}, 8 (1999)
  [Erratum-ibid.\  B {\bf 474}, 235 (2000)].

\bibitem{epj}
  S.~Scopetta and V.~Vento,
  Eur.\ Phys.\ J.\  A {\bf 16}, 527 (2003).


\bibitem{bpt}
  S.~Boffi, B.~Pasquini and M.~Traini,
  Nucl.\ Phys.\  B {\bf 649}, 243 (2003).


\bibitem{ruju}
  A.~De Rujula, H.~Georgi and S.~L.~Glashow,
  Phys.\ Rev.\  D {\bf 12}, 147 (1975).

\bibitem{ik}
  N.~Isgur and G.~Karl,
  Phys.\ Rev.\  D {\bf 18}, 4187 (1978);
  Phys.\ Rev.\  D {\bf 19}, 2653 (1979).
  [Erratum-ibid.\  D {\bf 23}, 817 (1981)].

\bibitem{Burkardt:2004ur}
  M.~Burkardt,
  Phys.\ Rev.\  D {\bf 69} (2004) 091501;
  Phys.\ Rev.\  D {\bf 69} (2004) 057501.

\bibitem{bro} S.J. Brodsky and S. Gardner, Phys. Lett. B 643, 22 (2006).  
\bibitem{ceb} 
E-06-010 Proposal to JLab-PAC29, J.-P. Chen and J.-C. Peng Spokespersons;
E-06-011 Proposal to JLab-PAC29, E. Cisbani and H. Gao Spokespersons. 
\bibitem{mio} S.~Scopetta, Phys. Rev. D {\bf 75}, 054005 (2007).
\bibitem{pisa} A. Kievsky, M. Viviani, and S. Rosati,
Nucl. Phys. A 577, 511 (1994).
\bibitem{av18} R.B. Wiringa, V.G.J. Stocks, and R. Schiavilla,
Phys. Rev. C 51, 38 (1995).
\bibitem{over} A. Kievsky, E. Pace, G. Salm\`e, and M. Viviani,
Phys. Rev. C 56, 64 (1997); E. Pace, G. Salm\`e, S. Scopetta, and A. Kievsky,
Phys. Rev. C 64, 055203 (2001).
\bibitem{model} 
D. Amrath, A. Bacchetta, and A. Metz, Phys. Rev. D 71, 
114018 (2005).
\bibitem{old} C. Ciofi degli Atti, S. Scopetta, E. Pace and G. Salm\`e,
Phys. Rev. C48, 968 (1993).
\end{thebibliography}

%


\end{document}